\documentstyle[aps,prb]{revtex}

\begin{document}
\draft
\title{On the conformational structure of a stiff homopolymer}
\author{Yu.A.~Kuznetsov, E.G.~Timoshenko\thanks{Corresponding author. 
Internet: http://darkstar.ucd.ie; 
E-mail: Edward.Timoshenko@ucd.ie}
}
\address{
Theory and Computation Group,
Department of Chemistry,\\ University College Dublin,
Belfield, Dublin 4, Ireland}
\date{\today}
\maketitle

\begin{abstract}
In this paper we complete the study of the phase diagram and conformational 
states of a stiff homopolymer.
It is known that folding of a sufficiently stiff
chain results in formation of a torus.
We find that the phase diagram obtained from the Gaussian variational
treatment actually contains not one, but several
distinct toroidal states distinguished by the winding number.
Such states are separated by first order transition curves terminating in
critical points at low values of the stiffness.
These findings are further supported by off--lattice Monte Carlo simulation. 
Moreover, the simulation shows that
the kinetics of folding of a stiff chain passes through various
metastable states corresponding to hairpin conformations with 
abrupt U-turns.
\end{abstract}


\section{Introduction}
\label{sec:intro}

Conformational transitions of semi--flexible polymers have
been of considerable interest for analytical studies
\cite{Harnau-95,Ganazzoli-95,GscTor,Maggs,Vasil-97} 
and computer simulations \cite{FrenkelStiff,Alan-MC-Mes,Binder-98} recently.
Polymers can possess different degree of flexibility
and their persistent length $\lambda$ is one of the key
parameters that determine the conformation
\cite{Polymer-books,Cloizeaux-book}.
For example, for polystyrene
$\lambda \simeq 1.4$ nm, which corresponds to about 5 chain bonds,
whereas for the double helix DNA $\lambda \simeq 50$ nm, i.e. about
150 base pairs.

The equilibrium \cite{Polymer-books,Cloizeaux-book} and kinetics of folding 
\cite{GscKinet,GscHomKin,Pitard} of a flexible 
homopolymer ($\lambda=0$) are relatively well understood
at present. It is believed that in a wide range of interaction
parameters the equilibrium collapse transition is continuous.
As the persistent length increases the extended coil becomes
larger with the swelling exponent changing smoothly from that of
the Flory coil $\nu=3/5$ to that of a rigid rod $\nu=1$.
The collapse transition of a semi--flexible homopolymer
on going from the good to the poor
solvent regime remains continuous at first, but starting from
some critical value of $\lambda$ becomes discontinuous 
\cite{Polymer-books,GscTor}.
Such a change of the nature of the transition actually
reflects a profound restructuring of the collapsed globule.
Thus, instead of forming a nearly spherical liquid--like globule
a sufficiently stiff chain has no point of easy bending so that
it wraps around itself forming a {\it torus}.
Such structures have been observed a number of times 
experimentally for DNA molecules \cite{DNAexper}
and were studied theoretically
from different points of view \cite{DNAtheor} starting from a pioneering
work \cite{TorusFirst}.

In works Refs. \onlinecite{Vasil-97,GscTor} it has been noted that for a
given degree of polymerisation the toroidal conformation exists
in a region starting from some value
of stiffness $\lambda$ and in a limited interval of
the solvent quality. In Ref. \onlinecite{GscTor} based
on the Gaussian self--consistent (GSC) method we have studied
the equilibrium phase diagram and kinetics of conformational
transitions after various quenches.  Importantly, both transitions
coil--to--torus and torus--to--globule are discontinuous,
and thus there are associated regions of metastability.
This results in a rather complex kinetic picture of expansion or folding, 
essentially dependent on the quench depth.
In that work we have also mentioned that there seemed  to be
some additional minima of the free energy, the study of which has
been deferred until the current paper.

In Ref. \onlinecite{Conf-Tra} we have shown that at equilibrium
the GSC treatment precisely reduces to the Gibbs--Bogoliubov
variational method with a generic quadratic trial Hamiltonian.
However, in the extended variational space care should be taken
while finding the true free energy minima as these seem to be sensitive
to the limitations of the underlying polymer model.
The current model with a virial--type expansion
is believed to have problems at high densities.
Thus we shall re--examine the phase diagram of a semi--flexible
chain here in a more systematic and accurate manner.

Despite a relative ease for analytical theories to obtain
the toroidal conformation, computer simulations have been less
successful so far. In Ref. \onlinecite{Alan-MC-Mes} an attempt was made
to include the bending energy into Monte Carlo simulation in
a lattice model of Ref. \onlinecite{CoplmMonte}. Although a number
of metastable states corresponding to conformations with hairpin
and crystalline conformations were observed, the true equilibrium
state corresponding to the torus was not possible to obtain.
These difficulties are due to a number of circumstances.
First, a toroidal conformation appears only for sufficiently
long and stiff chains. However, the relaxation times for these
become enormous and the equilibrium state is hard to reach
during a limited simulation time.
Second, the lattice itself introduces a number of unfortunate
artefacts. This is because the rotational symmetry is broken, so
that the chain segments can lie only along a number of allowed
directions, and thus weak bendings are simply impossible.
So, the chain forms conformations with long straight sections
which then possess a U-turn, resembling a hairpin. Such abrupt
turns, however, produce a considerable energetic penalty,
so that the corresponding states have a higher free energy
than the true minimum.
Depending on the particular type of the lattice model used
the true minimum may still be a mis--shapen torus, or
if the smallest and only bendings are 90 degrees, one would expect
instead a kind of a solid--like crystalline ordering 
\cite{ThreeBodyLat,FrenkelStiff}.
Even though a kind of torus may be possible in the model
with links along 2-D and 3-D diagonals \cite{CoplmMonte}, 
such a state would be extremely hard to reach due to the lack of collective
moves and a huge number of metastable minima in which the system
would tend to be trapped.
Interestingly, in a recent paper Ref.~\onlinecite{Binder-98} a simulation
in the bond fluctuation model, which is intermediate between the
lattice and continuous space models, has exhibited a toroidal,
though imperfect, structure (see e.g.~Fig.~7).

One attractive possibility is to use Langevin or Monte Carlo
simulations in continuous space instead, although this
would require much longer computations.
Here we shall use a
fairly standard Monte Carlo off--lattice technique for a ring
homopolymer chain.

Thus, the main objective of this work is by using both the Gaussian
variational and Monte Carlo techniques to obtain a more
accurate phase diagram of the homopolymer in terms of the stiffness
and the solvent quality and to elucidate the conformational 
structure of corresponding thermodynamically stable as well as possible 
metastable states.

\section{Techniques}
\label{sec:model}

\subsection{Gaussian variational method}
\label{subsec:gsc}

The Gaussian variational method is based on minimising the
Gibbs--Bogoliubov trial free energy,
${\cal A} \equiv {\cal E} - T {\cal S}$, with respect to
the full set of variational parameters.
Here these are the mean--squared distances,
$D_{mm'} \equiv (1/3) \langle ({\bf X}_m - {\bf X}_{m'})^2 \rangle$,
between monomers number $m$ and $m'$
($m,m'=0, \ldots N-1$, where $N$ is the degree of polymerisation).
For a ring homopolymer the matrix $D_{mm'}$
is translationally invariant along the chain: $D_{mm'}\equiv D_k$,
where $k=|m-m'|$.

Using the de~Gennes--des~Cloizeaux--Edwards bead--and--spring model
\cite{Polymer-books,Cloizeaux-book}
of the homopolymer with the volume interactions represented by the
virial--type expansion, one can obtain the following
entropic and the energetic contributions \cite{GscTor,Conf-Tra},
\begin{eqnarray}
{\cal S} & = & \frac{3 k_B}{2} \sum_{q=1}^{N-1} \log {\cal F}_q, \qquad
{\cal F}_q = -\frac{1}{2N} \sum_{k=1}^{N-1} 
\cos\left(\frac{2\pi kq}{N}\right) D_k, \qquad
R_g^2 \equiv \sum_{q=1}^{N-1}{\cal F}_q,
\label{gsc:Entropy} \\
{\cal E} & = & \frac{3 N\,k_B T}{2\,l^2} D_1
   + \frac{3 N\,k_B T \lambda}{2 l^3} (4 D_1 - D_2) 
   + \frac{u^{(2)} N}{(2\pi)^{3/2}} \sum_{k=1}^{N-1} D_k^{-3/2} +
 \frac{3 u^{(3)} N}{(2\pi)^3} \sum_{k=1}^{N-1} D_k^{-3} \nonumber \\
&& + \frac{u^{(3)} N}{(2\pi)^{3}} \sum_{k_1 \not= k_2 = 1}^{N-1} \left(
   \frac{1}{2}(D_{k_1} D_{k_2} + D_{k_1} D_{k_1-k_2} + D_{k_2} D_{k_1-k_2})
 - \frac{1}{4}(D_{k_1}^2 + D_{k_2}^2 + D_{k_1-k_2}^2)
     \right)^{-3/2}, \label{gsc:meanE}
\end{eqnarray}
where we have also introduced the normal modes ${\cal F}_q$ and the
mean squared radius of gyration of the chain $R_g^2$.

Analogously to Eqs.~(8-11) of Ref. \onlinecite{GscTor}
the first term  in Eq. (\ref{gsc:meanE}) is the elastic energy
of springs and the second term is the bending
energy, with $l$ and $\lambda$ being the statistical 
and the persistent lengths of the chain respectively.
Note that these formulas correspond precisely to the
Kratky--Porod model of taking the stiffness into account by
adding the integral along the chain of the squared curvature 
\cite{Kratky,Harris}.

In Refs. \onlinecite{Conf-Tra,Nuovo-Qz} we have discussed that
the application of a virial--type expansion is flawed for the dense
globular state. Namely, the model with the two-- and three--body terms 
\cite{Three-Body-Model} in the variational treatment is found to
possess a number of pathological infinitely deep free energy minima.
This implies that for a sufficiently high density the three--body term
is unable to cope with the increasingly strong two--body attraction.
Introduction of a thermodynamically subdominant term such as
the 4th term in Eq. (\ref{gsc:meanE}), ${\cal E}_{si}$ (see Appendix for
more details),
or the so--called `thickness' term in Ref.
\onlinecite{Allegra-Thickness}, fixes the problem, and is also
shown to produce a negligibly weak correction in the repulsive and
ideal coil regimes. Although attempts to derive such a term from
the perturbation and renomalisation theory have been made (as e.g. in
the latter work), these partial fixes are fundamentally inconsistent.
Having convinced ourselves that for the purpose of the current work
the results from such a theory are in satisfactory agreement with
the numerical experiment, we shall accept this procedure here
and bear in mind its limitations.
Hopefully, a new non--Gaussian theory under development
by one of the authors \cite{EGT-nonGSC} may address this problem.
It fundamentally deals with true intermolecular interaction
potentials instead of ill--defined virial expansions, something which
the Gaussian variational theory can not avoid since the energy averaged
over a Gaussian trial distribution diverges for any singular potential
involving a hard--core part.

\subsection{Off--Lattice Monte Carlo Simulation}
\label{subsec:cmc}

Since the lattice Monte Carlo model \cite{CoplmMonte,Alan-MC-Mes}
is not suitable for study of the effects of chain
stiffness for several reasons, we have carried out simulation
in continuous space instead.
The disadvantages of the lattice model include the
rotational anisotropy and tendency for condensed phases
to form some kind of crystalline structures on the lattice 
\cite{ThreeBodyLat}.
Thus, to produce toroidal states on a lattice for
comparatively short polymer chains is rather difficult.
Another disadvantage of the lattice model is that the
persistence of the polymer chain reduces dramatically the
acceptance ratio of the local monomer moves and, thus, other
more sophisticated types of moves such as shifts and rotations of chain
segments as a whole are needed.

The model is implemented for a single homopolymer consisting of $N$
monomers connected by springs in a ring, which
additionally interact with each other via a pair--wise short ranged
spherically symmetric potential,
\begin{equation}
H = \frac{k_B T}{2l^2}         \sum_m ({\bf X}_m - {\bf X}_{m-1})^2
  + \frac{k_B T \lambda}{2l^3} \sum_m ({\bf X}_{m+1}
                     + {\bf X}_{m-1} - 2{\bf X}_m)^2
  + \frac{1}{2} \sum_{m\not= m'} V (|{\bf X}_m - {\bf X}_{m'}|).
\label{cmc:hamil}
\end{equation}
Unlike the GSC theory, where one has to introduce a
virial--type expansion representing the pair--wise potential,
here we use the two--body interaction potential explicitly,
\begin{equation}
V(r) = \left\{
\begin{array}{ll}
+\infty & \quad\mbox{for\ } r < d \\
V_0 \left( \left( \frac{d}{r}\right)^{12}
- \left( \frac{d}{r} \right)^{6} \right) & \quad\mbox{for\ } r > d
\end{array}
\right.. \label{cmc:V}
\end{equation}
Thus, monomers are represented by hard spheres of the diameter $d$,
with a weak short ranged Lennard--Jones attraction of characteristic
strength $V_0$. During simulation 
we change the strength of the two--body attraction $V_0$,
which can be viewed as basically the ``inverse temperature'',
rather than changing the temperature $T$ itself.

The Monte Carlo updates scheme is based on the Metropolis
algorithm with local monomer moves.
The new coordinate of a monomer can be sought as,
$q^{new} = q^{old} + r_{\Delta}$,
where $q$ stands for $x$, $y$ and $z$ spatial projections and
$r_{\Delta}$ is a random number uniformly distributed in the
interval $[-\Delta,\,\Delta]$. Here $\Delta$ is some additional
parameter of the Monte Carlo scheme, which, in a sense,
characterises the timescale involved in the Monte Carlo sweep (MCS),
the latter being defined as $N$ attempted Monte Carlo steps.

In both models we work in the system of units such that $l=1$ and
$k_B T =1$. Additionally we fix the third virial coefficient in the
variational method, $u^{(3)} = 10$, and the hard--core diameter, $d = 1$,
in Eq.~(\ref{cmc:V}).

\section{Equilibrium Phase Diagram from the variational method}
\label{sec:eql}

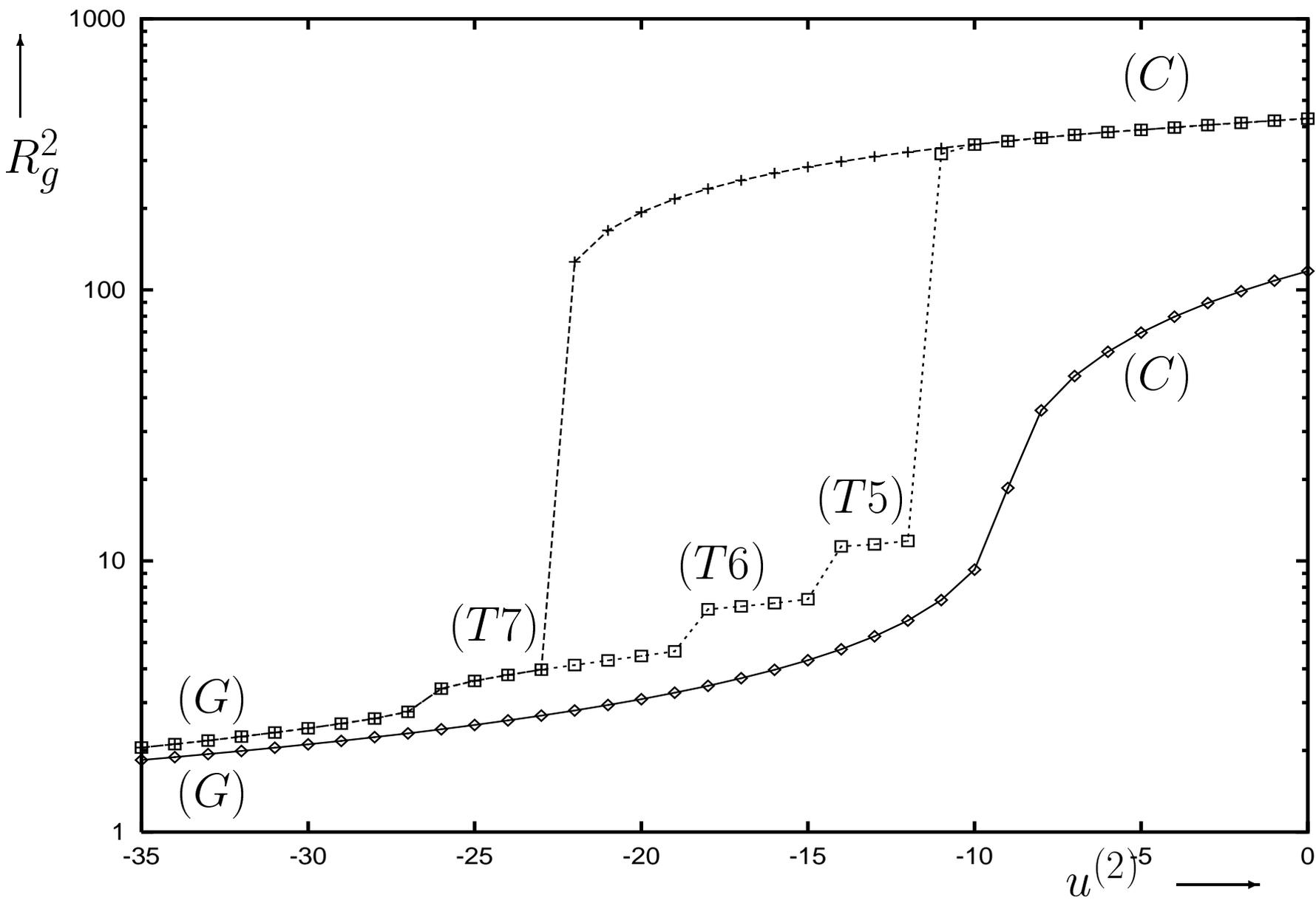
\begin{figure}
\caption{ \label{fig:t2_rg2u}
Plot of the mean squared radius of gyration, $R_g^2$, vs the second
virial coefficient, $u^{(2)}$, upon a quasistatic change of the latter
for ring homopolymer with $N=100$.
The solid line (diamonds) corresponds to the value of the 
stiffness parameter $\lambda = 5$, for which the collapse transition 
is second order.
The dashed lines correspond to $\lambda = 25$ and are obtained by
going from positive values of $u^{(2)}$ to negative (pluses) and
backwards (quadrangles).
Here labels $(C)$, $(G)$, and $(T5)$-$(T7)$ correspond to the coil,
the globule and various toroidal states respectively.
}
\end{figure}

First, let us consider the system behaviour upon a quasistatic
change of the interaction parameters. In Fig.~\ref{fig:t2_rg2u} we
present the plot of the mean squared radius of gyration,
$R_g^2$, versus the second virial coefficient, $u^{(2)}$,
at a fixed stiffness parameter, $\lambda$.
The regime of repulsion and comparatively weak attraction
in the right--hand--side of the figure
corresponds to the extended coil conformation of the polymer
with a large radius of gyration scaling as $R_g \sim N^{\nu_{coil}}$,
where the exponent $\nu_{coil}$ is close to the Flory value
$\nu_F = 3/5$ for a flexible chain, $\lambda=0$, becomes a
rigid rod exponent $\nu_{rod}=1$ for a very stiff chain,
with a cross--over in between. Since increasing the stiffness leads to a 
stronger effective repulsion between monomers, the extended phase expands 
to the region of the negative second virial coefficient for higher values
of the stiffness parameter.

For comparatively small
values of $\lambda$ the plot of the radius of gyration (solid line
and diamonds in Fig.~\ref{fig:t2_rg2u}) is quite similar to that
of the flexible homopolymer at the equilibrium coil--to--globule
transition (see e.g.~Fig.~1 in Ref.~\onlinecite{GscHomKin}),
which is second order.
However, at higher values of the stiffness parameter
the collapse transition becomes first order \cite{Footnote1}.
In this case, after the system has been quasistatically quenched to
the region of a higher monomer attraction (line denoted by pluses in
Fig.~\ref{fig:t2_rg2u}), the local minimum corresponding
to the coil suddenly disappears becoming an inflexion point
somewhere in the interval, $-23 < u^{(2)} < -22$, and the system
passes to another free energy minimum with a much smaller value of
the radius of gyration.
Similarly, upon changing $u^{(2)}$ quasistatically towards monomer
repulsion (line denoted by quadrangles in Fig.~\ref{fig:t2_rg2u}),
the free energy minimum disappears in the interval
$-12 < u^{(2)} < -11$, and the system transforms into the coil state.
If at least two minima of the free energy can coexist in some
interval of the interaction parameters, the transition point
is defined by the condition that
the current minimum of the free energy ${\cal A}$ becomes the deepest one.
Observables such as  the mean energy, ${\cal E}$,
the mean squared radius of gyration, $R_g^2$, and the mean squared distances
between monomers, $D_{mm'}$, experience a discontinuous jump
at such a transition.

It is important to note that upon a quasistatic change
of the second virial coefficient towards repulsion
(line denoted by quadrangles in Fig.~\ref{fig:t2_rg2u})
the mean squared radius of gyration increases in a
three--step--like fashion before the homopolymer expands to the
coil. This is a manifestation of some additional
condensed phases, which we have denoted by labels
$(T5)$, $(T6)$ and $(T7)$. To understand the distinction
between these phases and the conventional globule let us
compare the monomer--monomer mean squared distances, $D_k$,
in these phases. These are exhibited in Figs.~\ref{fig:t2_dmm}a, b.
As we have discussed earlier \cite{GscKinet,GscHomKin}
for the state of the extended coil this function
monotonically increases on the half--period of the chain.
The situation remains similar for the coil of a
stiff homopolymer (line denoted by diamonds in Fig.~\ref{fig:t2_dmm}a).

However, the function $D_k$ is more sensitive to the stiffness in the 
state of the globule,
especially at small values of the chain index $k$ (compare 
the line denoted by pluses with the solid line in
Fig.~\ref{fig:t2_dmm}a).
At small values of $k$ the function is nearly parabolic, 
$D_k \sim |k|^2$, i.e. the chain represents almost a rigid rod, reaching
a maximum at some value of the chain index $k^*$.
In some intermediate range of the chain index, $0 < k \lesssim 6k^*$,
one can see about 2-3 oscillations in the function $D_k$,
with the amplitude decreasing quickly to stabilise at some level. 
At higher values of the chain
index towards half of the chain the function remains constant.
Thus, we can conclude that for small chain distances
the structure of the globule of a fairly stiff polymer
is quite different from that of the flexible chain.
This is easy to understand.
For a semi--flexible polymer chain segments in the globule 
are locally straightened on a characteristic scale related to $\lambda$.
As long as this scale is considerably smaller than the globule size
the shape of $D_k$ remains flat as for the flexible chain.
When this scale becomes comparable
to the globule size, a few oscillations appear in the mean squared distances.

Transition from the conventional globule to the phase labelled
as $(T7)$ is accompanied by a spectacular change of the
function $D_k$
(see line denoted by quadrangles in Fig.~\ref{fig:t2_dmm}b).
In this phase the function strongly oscillates with the amplitude
decreasing rather slowly towards half of the chain. 
The ratio of the value of $D_k$ in a maximum to that in a minimum is 
about 5-6 near the middle of the chain.
The designation of the phases is done according to the number of 
oscillations:
in phase $(T7)$ ther are 7 oscillations, in phase $(T6)$ there are 6 
oscillations (line denoted by pluses
in Fig.~\ref{fig:t2_dmm}b), and in phase
$(T5)$ there are 5 oscillations (line denoted by diamonds).
We also note that
the smaller the number of oscillations the higher is
the ratio of the value of $D_k$ in a maximum to that in a minimum.

We claim that the phases $(Tn)$ correspond to the toroidal
conformation with the number of windings ${\cal N}_w=n$.
The chain index $k^*$, where the function $D_k$ reaches its first
maximum, is equal to the number of monomers forming 
half--period of the first winding starting from the zeroth monomer.
Therefore, $D_{max} = D_{k^*}$  may be interpreted roughly
as the mean squared {\it external} diameter of the torus.
By moving from the monomer number $k^*$ to $2\,k^*$
the first winding is completed. However, because of the excluded
volume interaction the chain cannot return to the same coordinate,
giving rise in the average to the value $D_{min} = D_{2k^*}$,
which may be considered as the mean squared {\it internal}
diameter of the torus.
The winding number ${\cal N}_w$ is thus precisely the number of
oscillations in $D_k$.
The physical reason for a torus is clear --- a persistent chain
has no desire to bend, so it tends to have as large a radius of
curvature as possible, however, two--body attraction tends to
keep quite close packing of the chain.

A quasistatic increase of the repulsion (see line denoted by
quadrangles in Fig.~\ref{fig:t2_rg2u}) results in transformation of
the conventional globule to the toroidal globule with ${\cal N}_w=7$
windings in the interval, $-27 < u^{(2)} < -26$. This is a rather
weak discontinuous transition.
It occurs when the characteristic scale of straightened segments reaches
the size of the globule,
and a hole in the centre of the globule is formed.
Transitions between various toroidal states are also first
order, but much stronger since they are accompanied by a global
restructuring of the polymer conformation.

In Fig.~\ref{fig:t2_rg2r} we present the plots of the
mean squared radius of gyration, $R_g^2$, upon a quasistatic
change of the stiffness parameter $\lambda$ for
different fixed negative values of the second virial
coefficient, $u^{(2)}$, corresponding to the globule at $\lambda=0$.
The radius of gyration increases monotonically during this change.
The most significant changes occur in the region of rather small
stiffness parameter, $0 < \lambda < 1/2$, and in the region where
the conventional globule transforms to the toroidal state.
The second change is associated with a weak first order transition. 
Note that the final toroidal state depends on the value of the second virial
coefficient. At a weaker attraction the globule is transformed
to a toroidal state with a smaller winding number.
However, the transition from the coil to the toroidal phases
upon a quasistatic change of $u^{(2)}$ (line denoted by pluses in
Fig.~\ref{fig:t2_rg2u}) is quite difficult due to a large
potential barrier, which makes the coil a metastable state
in a large region of the phase diagram.

In Fig.~\ref{fig:t2_phd} we present the resulting phase diagram of the
stiff homopolymer in terms of the second virial coefficient,
$u^{(2)}$, and the stiffness parameter, $\lambda$.
It contains phases of the coil, where monomer attraction is insufficient
to form compact states, the globule in the region of
either a low stiffness or a strong monomer attraction, and a number of
toroidal phases characterised by distinct winding number ${\cal N}_w$.
As we have already discussed above the collapse transition changes
its behaviour from continuous to discontinuous
starting from some value of the stiffness.
The globule of a semi--flexible polymer is different in the local structure
from that of the flexible homopolymer, although global
scaling characteristics are the same for both cases. The toroidal
phases lie in the intermediate region in $u^{(2)}$ starting from
some critical values of the stiffness parameter.
Some of such states with ${\cal N}_w=2,3,4$ are always metastable, while
some with ${\cal N}_w=5,6,7$ can become thermodynamically stable.

For a large fixed value of the stiffness parameter the number of
toroidal phases increases approximately linearly with the degree
of polymerisation, $N$. For example, the maximal winding numbers
at $\lambda = 25$ for polymers with the degree of polymerisation
$N = 50$, $100$, $150$ and $200$ are equal to:
${\cal N}_w = 4$, $7$,  $10$  and $13$ respectively.

\section{Results from Monte Carlo simulation}
\label{sec:mc}

In series of pictures in Fig.~\ref{fig:t2_CMC_eql} we exhibit
typical conformations in various phases from the off--lattice
Monte Carlo simulation.
Fig.~\ref{fig:t2_CMC_eql}a corresponds to the extended conformation,
which for large values of the stiffness parameter takes a form
close to a ring. In Fig.~\ref{fig:t2_CMC_eql}b we draw the
backbone of a typical globular conformation for the values of the
stiffness parameter not large enough to form a toroidal state.
The globule structure here is quite different from that of a flexible
homopolymer in Fig.~\ref{fig:t2_CMC_eql}c.
One can see that it consists of entangled loops of
a radius close to the size of the globule. The function of the
mean squared distances between monomers here possesses a typical
form presented by pluses in Fig.~\ref{fig:t2_dmm}a with
significant oscillations on small chain distances, which quickly
saturate due to varying number of monomers in each loop.
We should also note that the globule of a stiff homopolymer is
not quite spherical, but rather reminds an ellipsoid, either
flattened or elongated. Thus, we avoid calling it ``spherical
globule'' as we did in Ref. \onlinecite{GscTor}.
Increasing the stiffness parameter transforms such a globule to
the toroidal conformation exhibited in Fig.~\ref{fig:t2_CMC_eql}d.

It is important to note that to produce the globule state as in
Fig.~\ref{fig:t2_CMC_eql}b, or the toroidal state
as in Fig.~\ref{fig:t2_CMC_eql}d, in a Monte Carlo simulation it
is much easier to bring the system first to
the globule of the flexible homopolymer (Fig. \ref{fig:t2_CMC_eql}c),
and then to increase the stiffness parameter quasistatically.
If instead we change $V_0$ quasistatically at a fixed sufficiently 
large $\lambda$ reaching the equilibrium would be difficult.
First, as we have mentioned earlier, the region of the metastable coil
is rather wide for large values of the stiffness.
Second, the system may become trapped in a metastable state during such a
process. Polymer conformation in these states have a typical
form of a {\it hairpin} (see Fig. \ref{fig:t2_CMC_kin10}c).
Here the chain folds a few times
along a nearly straight line forming abrupt U-turns near the ends.
However, these ends contribute more significantly to the
bending energy than a uniform slow bending. Thus, such
conformations possess a higher energy and hence are metastable.

Let us consider in more detail the kinetic process of 
pairpin formation after an
instantaneous change of the two--body attraction, $V_0$, starting
from the ring--like conformation as in Fig.~\ref{fig:t2_CMC_eql}a.
Due to the stiffness the initial ring conformation remains stable
for some time.
However, due to thermal fluctuations
some distant parts of the chain would meet each other occasionally,
so that the chain acquires a shape of the digit `8' as
in Fig. \ref{fig:t2_CMC_kin10}a.
If monomers are attractive enough, parts of the chain would
align along each other starting from the centre towards ends 
(see Fig. \ref{fig:t2_CMC_kin10}b).
The process can repeat itself if the persistent length
is shorter than half of the chain.
Without performing quite sophisticated collective movements
of the chain segments it is virtually
impossible to proceed further towards the fully collapsed state.
Snapshot in Fig. \ref{fig:t2_CMC_kin10}c corresponds to $10^7$ MCSs
and the system is still trapped in the same hairpin state.

In series of pictures in Fig.~\ref{fig:t2_CMC_kin5} we exhibit
polymer conformations during kinetics of folding for a smaller
value of the stiffness parameter, $\lambda=5$. After a long
evolution during which the ring remains practically unchanged
some loops of a size comparable to the persistent length are formed
(see Fig. \ref{fig:t2_CMC_kin5}a), which continue to grow
by picking up the slack of the chain and thus forcing a few more 
loops to form. Then a kind of a star--like structure including a few hairpins
is produced as in Fig. \ref{fig:t2_CMC_kin5}b. These hairpins fold
onto each other producing a {\it sausage}--like object
(see Fig. \ref{fig:t2_CMC_kin5}c). Further rather slow kinetic process
involves refolding of the sausage which is accompanied by its broadening
and shortening (see Fig. \ref{fig:t2_CMC_kin5}d).
Thus, kinetics of folding leads to elongated sausage--like conformaions,
whereas a similar quasistatic change of the stiffness would tend to produce
a flattened rather than elongated globule (see Fig. \ref{fig:t2_CMC_eql}b).

Finally, let us note that the hairpin conformations have not been obtained
by the GSC method neither at equilibrium,
nor even as intermediate kinetic states, although they are present as
local free energy minima.
This is probably related to that in the Gaussian method the
monomer--monomer correlation functions are represented by
only one parameter characterising the mean squared distances.
On the other hand, the lack of collective moves in the Monte
Carlo scheme quite likely overestimates the stability of hairpin 
conformations as metastable or kinetically arrested states.

\section{Conclusion and Discussion}
\label{sec:concl}

In this paper we have completed the study of the phase
diagram of a stiff homopolymer based on the Gaussian variational
method and have also performed some equilibrium and kinetic
Monte Carlo simulations in continuous space.
Compared to the previous work Ref. \onlinecite{GscTor} we
have shown here that the region corresponding to the toroidal
globule actually consists of a number of strips corresponding to
tori with different {\it winding numbers}. The transition curves separating
such states from each other, as well as from the coil and the spherical
globule, all correspond to first order transitions and terminate in 
critical points. For a given sufficiently large degree of
polymerisation $N$ there exist a certain number of different toroidal
states, which grows with $N$ approximately linearly. The distinction
between such toroidal states is clearly visible in the function of the 
mean squared distances between monomers, which shows precisely
the number of oscillations equal to the winding number.
The existence of the toroidal states has been also confirmed by
off--lattice Monte Carlo simulation for the similar model of
local stiffness. In addition, hairpin conformations with abrupt U-turns
corresponding to metastable states have been observed.
These also appear as nonequilibrium intermediaries during kinetics
of folding.

We should emphasise that the results in the previous work Ref.
\onlinecite{GscTor}  at sufficiently strong monomer attractions
do suffer somewhat from the artefacts of the
de~Gennes--des~Cloizeaux--Edwards bead--and--spring model
based on the virial expansion. 
Here, by including the ${\cal E}_{si}$-term 
introduced in Ref. \onlinecite{Conf-Tra}, we have 
managed to fix the problem at a practical level, although its
fundamental resolution remains the matter of future work
\cite{EGT-nonGSC}. The changes in the results are as follows.
First, the region designated as `Torus' in Fig.~1 of 
Ref.\onlinecite{GscTor} expands and covers all of the region designated as 
`Spherical Globule' there
(i.e. curves III, III', III'' in Fig.~1 of Ref.
\onlinecite{GscTor} shift to the left of the axis $\lambda=0$). 
Indeed, a slightly oscillating behaviour
of $D_k$ is only related to the existence of locally straightened
sections of the chain, and does not necessarily correspond to a torus.
Second, the metastable states which we called T', T'' and so on in Fig.~6
of Ref. \onlinecite{GscTor} now become thermodynamically stable in some
regions of the phase diagram in Fig. \ref{fig:t2_phd}.
Oscillations of the function $D_k$ for the true toroidal states are
much stronger (see Fig. \ref{fig:t2_dmm}b) 
and this function never reaches a steady level. 
Most importantly, there are a few distinct toroidal states of a ring
polymer separated by first order transitions and distinguished by the winding
number. 
Nevertheless, the main conclusion of Ref.
\onlinecite{GscTor} that the toroidal conformation exists in a 
triangularly shaped region of the phase diagram, remains valid.
Most of the general conclusions about the stability of the toroidal
conformation and kinetics of conformational changes  
of Ref. \onlinecite{GscTor} remain unchanged too.

We would like to emphasise that the existence 
of the toroidal states is pretty much related to the choice of 
the bending energy as the square of the local curvature of the chain.
Indeed, such a choice is natural for representing the persistent flexibility
of polymers, which is due to rather small harmonic 
fluctuations in bending of the chain sections. This mechanism of
flexibility is dominant for many helical or rather stiff chains such
as DNA, for which the toroidal states have been observed experimentally
\cite{DNAexper}.

The rotational--isomeric flexibility is also very important for many polymers.
Such polymer molecules exhibit flexibility due to rotation around 
carbon--carbon and other bonds, as the minima of the torsional potential
corresponding to the {\it gauche} and {\it trans} configurations
have the difference in their depth of about $k_B T$.
For representing this mechanism of flexibility a model with discrete 
bending angles is more appropriate. Such models, however, possess 
crystalline solid--like states \cite{ThreeBodyLat,FrenkelStiff} 
instead of toroidal ones. Such a difference in conformational
states is genuine in our view and both types of structures are
observable experimentally depending on the particular polymer system.

Another important prerequisite for obtaining a toroidal state
even in a model with persistent flexibility mechanism is that
the processes of inter--chain aggregation do not occur,
otherwise more complicated self--assembled structures  may be formed.
For instance, aggregation of triple helix collagen molecules
leads to self--assembly of various fibrils.

\acknowledgments

The authors are grateful for most interesting discussions to
Professor A.Yu.~Grosberg and Professor K.A.~Dawson, and also to
Professor K.~Binder and Dr A.V.~Gorelov.
One of us (E.G.T.) acknowledges the support of the Enterprise
Ireland grants IC/1999/001 and SC/99/186.

\appendix

\section*{The self--interaction energy term}

Let us discuss in more detail the appearance of the last term, ${\cal E}_{si}$,
in Eq.~(\ref{gsc:meanE}),
which we call the self--interaction energy term.
Such a term has been introduced in Ref. \onlinecite{Conf-Tra} 
for heteropolymers,
for which the two--body interaction matrix,  $u^{(2)}_{mm'}$, is site
dependent.
Generally, one has to discard the singular terms with coinciding
indices in the virial expansion as we have done in Eq.~(\ref{gsc:meanE}).
It turns out, however, that the resulting free energy possesses some
pathological minima 
with singular free energy if at least one element of the
matrix $u^{(2)}_{mm'}$ becomes negative.
Indeed, let us
consider the interactions of just three monomers under the condition
that the mean squared distances from monomers '0' and '1' to '2' are
equal to each other, $D_{0,2} = D_{1,2} = D$. These interactions
produce the mean energy contribution, 
\begin{equation} \label{gsc:nons}
{\cal E}_3 = \frac{u^{(2)}_{0,2} + u^{(2)}_{1,2}}{(2\pi D)^{3/2}} +
             \frac{1}{(2\pi D_{0,1})^{3/2}} \left( u^{(2)}_{0,1} +
             \frac{6 u^{(3)}(2\pi)^{-3/2}}{(D - D_{0,1}/4)^{3/2}} \right)
\end{equation}
If $u^{(2)}_{0,1} < 0$ and monomer '2'
is placed away from monomers '0' and '1',
$D > D_{0,1}/4 + (6 u^{(3)}(2\pi)^{-3/2}/|u^{(2)}_{0,1}|)^{2/3}$,
obviously, in the limit $D_{0,1} \rightarrow 0$ the mean
energy possesses a singular minimum, ${\cal E}_3 \rightarrow -\infty$.
As for the free energy, the logarithmic divergence of the entropy could
not change the situation, thus ${\cal A} \rightarrow -\infty$ as well.
One can also show that the inclusion of more monomers in the chain,
or of higher than the three--body interactions, 
does not improve the situation, but produces
more and more of such pathological solutions.

The problem can be quite easily
remedied by using another prescription --- replacing the terms with
coinciding indices by the self--interaction terms \cite{Conf-Tra,Nuovo-Qz}.
Namely, we should add the following term,
\begin{equation} \label{gsc:Esi}
{\cal E}_{si} = c_3 u^{(3)}\sum_{m\not=m'}\biggl\langle
 \delta({\bf X}_{m}-{\bf X}_{m'}) \biggr\rangle^2 =
 c_3 \frac{u^{(3)}}{(2\pi)^3} N \sum_k D_k^{-3},
\end{equation}
where $c_3 = 3$ is a combinatorial factor related to the three possible
ways of having coinciding pairs of indices in a triple summation.
Obviously, the higher negative power of $D_k$ in Eq.~(\ref{gsc:Esi})
compared to the two--body term in Eq.~(\ref{gsc:meanE}) prevents
one monomer from falling onto another.

Interestingly enough, this problem is hidden for a ring homopolymer.
Indeed, due to the inverse symmetry
\cite{GscBlock}, we have the property that for any indices 
$m$, $m'$ the following
mean squared distances are equal: $D_{m,m'} = D_{m,2m-m'} = D_{2m'-m,m'}$.
This provides sufficient repulsion coming from the three--body term to preclude
any pathological solutions.
Nevertheless, without the self--interaction energy term the theory
possesses an unphysical behaviour: forcing two monomers very close 
to each other produces a rather weak repulsion of this pair 
from all very distant monomers. 
This repulsion comes from the three--body interaction 
and it is dominant at no matter how strong two--body
attraction between monomers.

This effect comes into play only for sufficiently large negative $u^{(2)}$.
In particular, it leads to a somewhat more convex than expected shape 
of the $D_k$ function for the globule of a flexible homopolymer.
Nevertheless, the kinetics of folding remains little affected by this
deficiency. Adding the self--interaction energy term improves the situation
and results in a flat shape of $D_k$ (see Ref. \onlinecite{Nuovo-Qz} 
for more detail).
For a stiff homopolymer there are even more pronounced problems
associated with the pathological three--body repulsion in the absence
of ${\cal E}_{si}$. 
So, from Fig. 4 of Ref. \onlinecite{GscTor} one can see that  there the
size of the spherical globule was larger than that of the toroidal
globule in the vicinity of the transition. 
Including the ${\cal E}_{si}$ term reverses this situation as we shall
see below. It also lowers the depths of the additional local free 
energy minima corresponding to phases such as
T' and T'' in Fig. 6 of Ref. \onlinecite{GscTor}.
Finally, the inclusion of the ${\cal E}_{si}$ term leads to a better
agreement with various results from Monte Carlo simulations for flexible
and stiff homopolymers.




\section*{Figure Captions}

\begin{figure}
\caption{ \label{fig:t2_dmm}
Plot of the mean  squared distances between monomers, $D_k$, vs the
chain index, $k$, for various conformational states of the homopolymer.
In Fig. \ref{fig:t2_dmm}a lines denoted by diamonds and pluses
correspond to the coil ($u^{(2)}=0$) and the globule ($u^{(2)}=-30$)
for $\lambda=25$. The solid line corresponds to the globule of flexible
homopolymer ($\lambda=0$ and $u^{(2)}=-25$).
In Fig. \ref{fig:t2_dmm}b lines denoted by diamonds,
pluses, and quadrangles correspond to the
toroidal states with the winding number ${\cal N}_w=5$, $6$ and $7$ 
($\lambda=25$, $u^{(2)} = -13$, $-17$ and $-25$) respectively.
}
\end{figure}

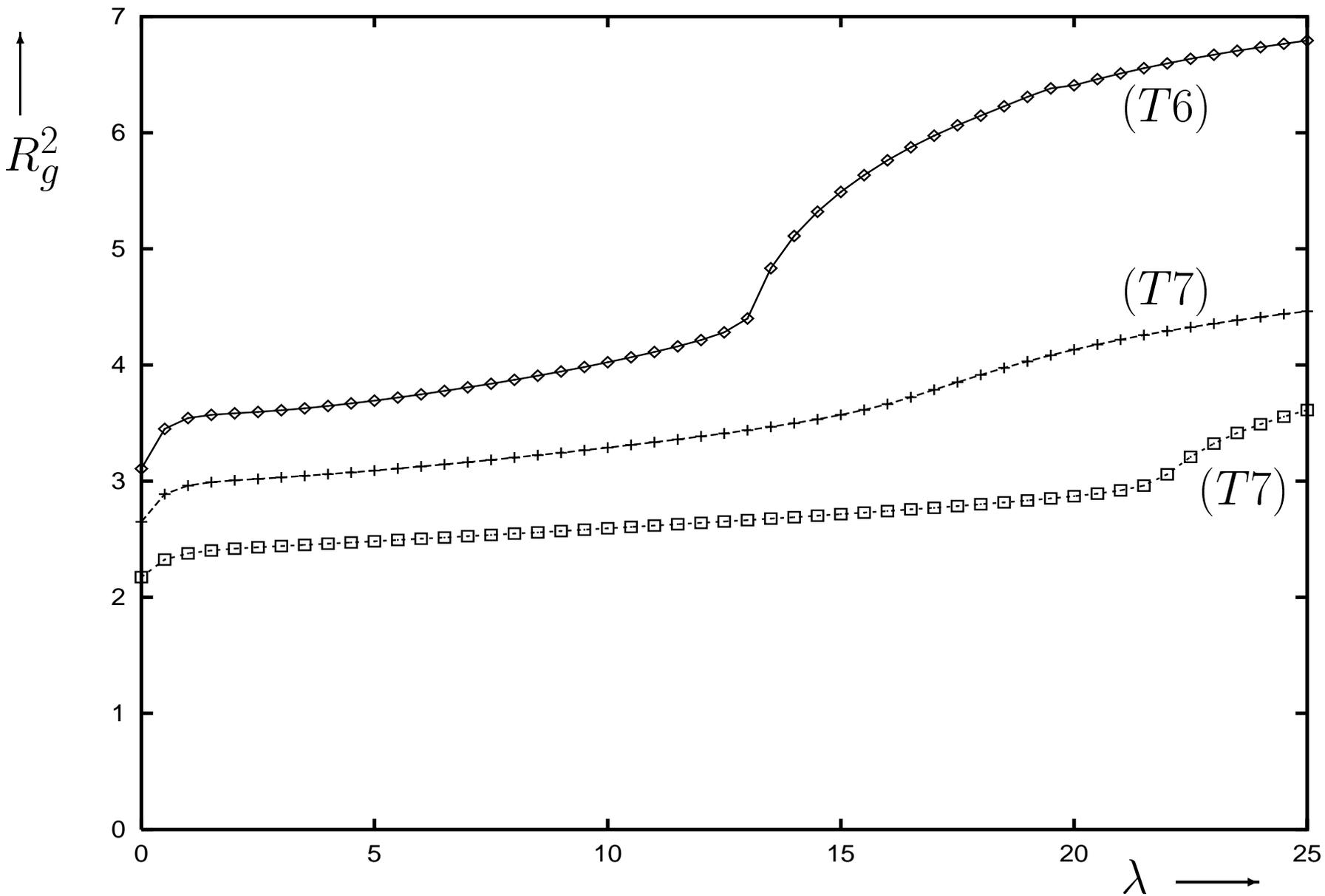
\begin{figure}
\caption{ \label{fig:t2_rg2r}
Plot of the mean  squared radius of gyration, $R_g^2$, vs the
stiffness parameter, $\lambda$.
Lines denoted by diamonds, pluses and quadrangles (from top to bottom)
correspond to  $u^{(2)} = -17$, $-20$ and $-25$ respectively.
}
\end{figure}

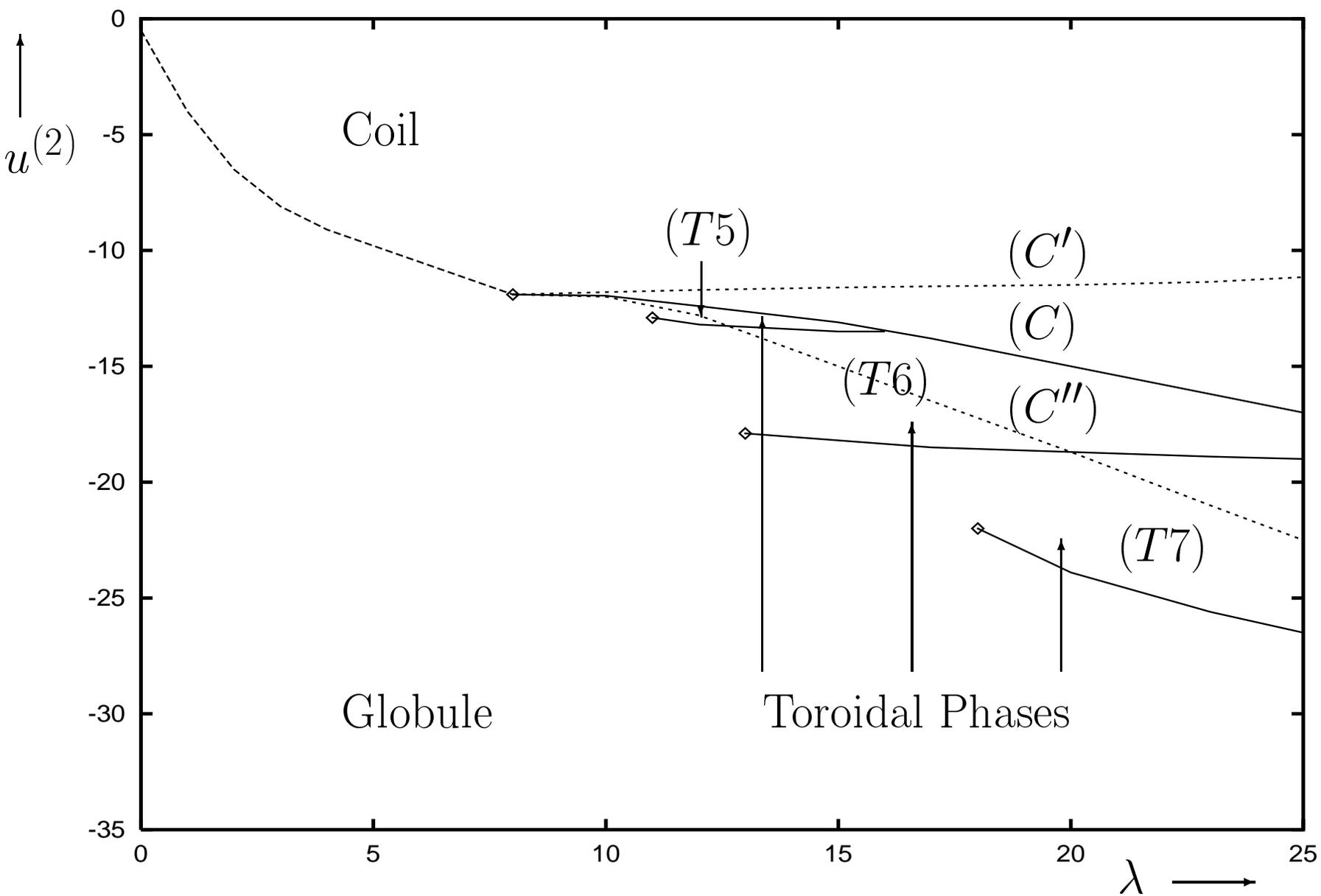
\begin{figure}
\caption{ \label{fig:t2_phd}
Phase diagram of the stiff homopolymer in terms of the stiffness
parameter, $\lambda$, and the second virial coefficient, $u^{(2)}$
for $N=100$.
Here $(C)$ denotes the transition curve between the coil and globular phases,
and curves $(C')$ and $(C'')$ are the metastability boundaries of that 
transition.
Labels $(T5)$-$(T7)$ correspond to toroidal states with
the winding number ${\cal N}_w=5$, $6$ and $7$ respectively. 
The metastability boundaries of these transitions
are not depicted. 
}
\end{figure}

\begin{figure}
\caption{ \label{fig:t2_CMC_eql}
Snapshots of typical equilibrium conformations of a 
homopolymer with the degree of polymerisation $N = 100$ 
from the off--lattice Monte Carlo simulation.
Figures (a)-(d) correspond to
the coil ($V_0 = 0$, $\lambda = 5$),
the globule of a semi--flexible homopolymer ($V_0 = 5$, $\lambda = 5$),
the globule of the flexible homopolymer ($V_0 = 5$, $\lambda = 0$),
and the toroidal globule ($V_0 = 5$, $\lambda = 15$).
}
\end{figure}

\begin{figure}
\caption{ \label{fig:t2_CMC_kin10}
Snapshots of typical conformations of a stiff ($\lambda=10$)
homopolymer with $N = 100$ during kinetics after 
after the quench from $V_0=0$ to $V_0=5$.
Figures (a)-(c) correspond to
the following moments in time:
$t=1.4\cdot 10^6$ MCS, $t=1.6 \cdot 10^6$ MCS, and $t=10^7$ MCS.
Here and in Fig. \ref{fig:t2_CMC_kin5}
the maximal attempted coordinate variance is $\Delta = 0.08$.
}
\end{figure}

\begin{figure}
\caption{ \label{fig:t2_CMC_kin5}
Snapshots of typical  conformations of a semi--flexible
($\lambda=5$) homopolymer 
with the degree of polymerisation $N = 100$ 
during kinetics after the same quench as
in Fig.~\ref{fig:t2_CMC_kin10}.
Figures (a)-(d) correspond to the following moments in time:
$t=350,000$ MCS, $t=800,000$ MCS, $t=10^6$ MCS, and $t=5\cdot 10^6$ MCS.
}
\end{figure}


\begin{references}
\bibitem{Footnote1}
We should emphasise here that the order of a transition is defined by us
according to how the main free energy minimum evolves upon a quasistatic 
movement in the parameter space.
Such a definition is perfectly well justified
even for finite--size systems, while the behaviour of some observables
may appear fairly smooth in both cases due to finite size effects.
This mean--field definition of the transition order manifests itself
kinetically in the existence of a nucleation lag for discontinuous
transitions. Unfortunately, in a computer simulation the determination
of the order of transitions and even more so of the full phase diagram
requires an enormous labour \cite{Binder-98}.
Practically, the boundary of a first order transition we determine here by the
condition of equality of the free energy values on both of the competing
minima, i.e. ${\cal A}_1 = {\cal A}_2$. For a second order transition
there is a finite transition width involved and so, to be precise,
we determine the transition point as the point of the maximal change
of a corresponding order parameter $X$ ($R_g^2$ in our case), i.e.
$dX/dT = 0$.
\end{references}
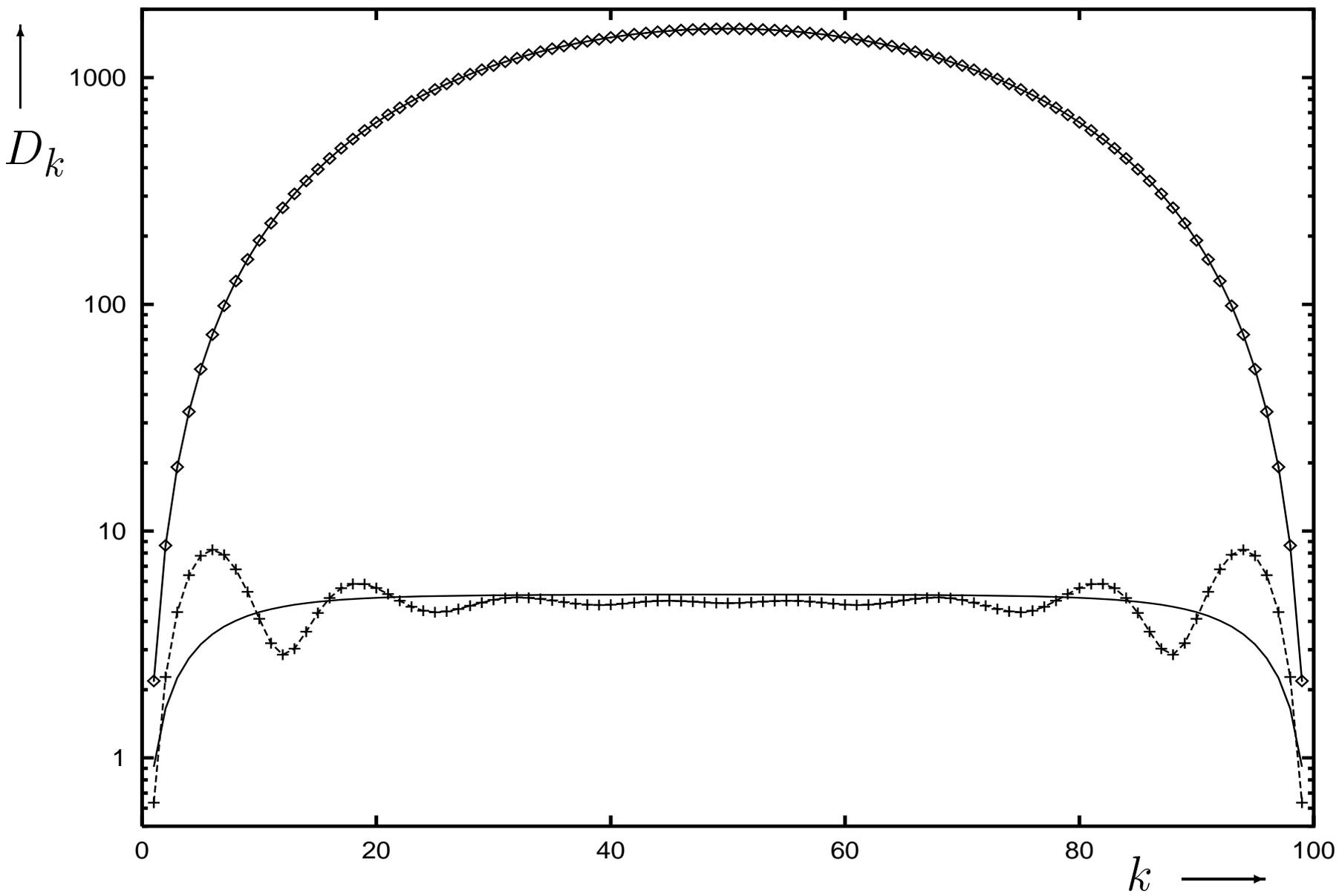
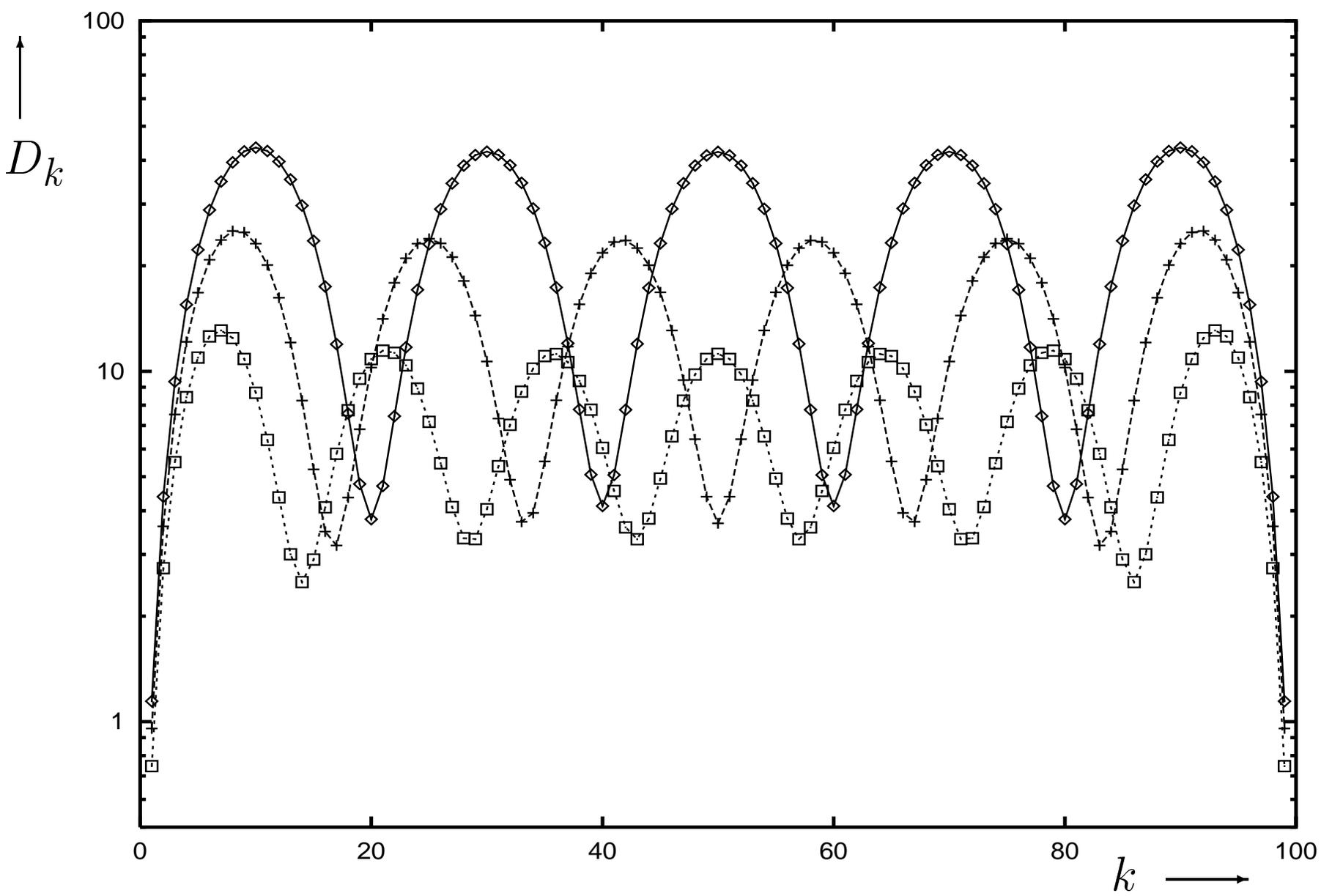
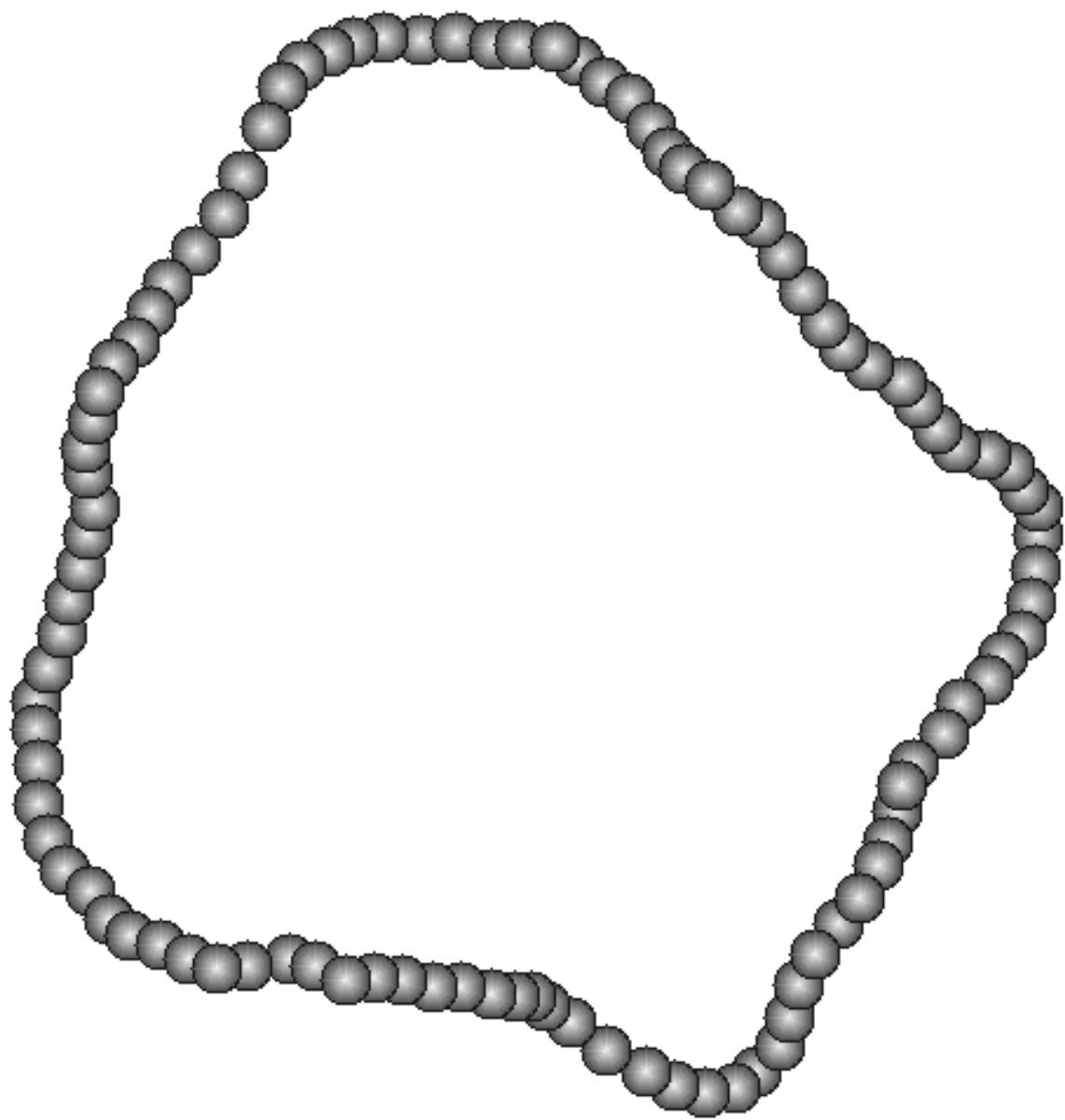
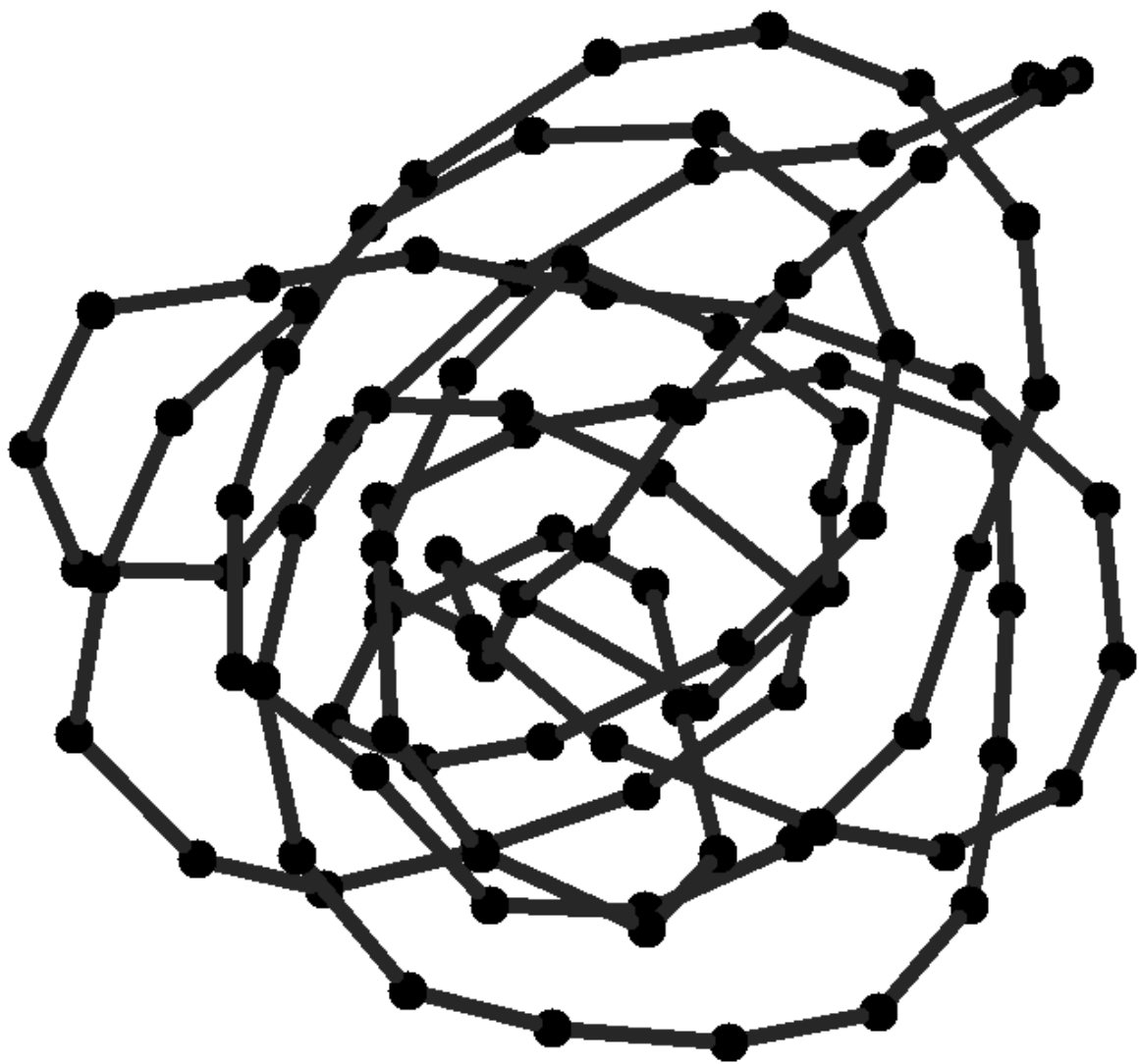
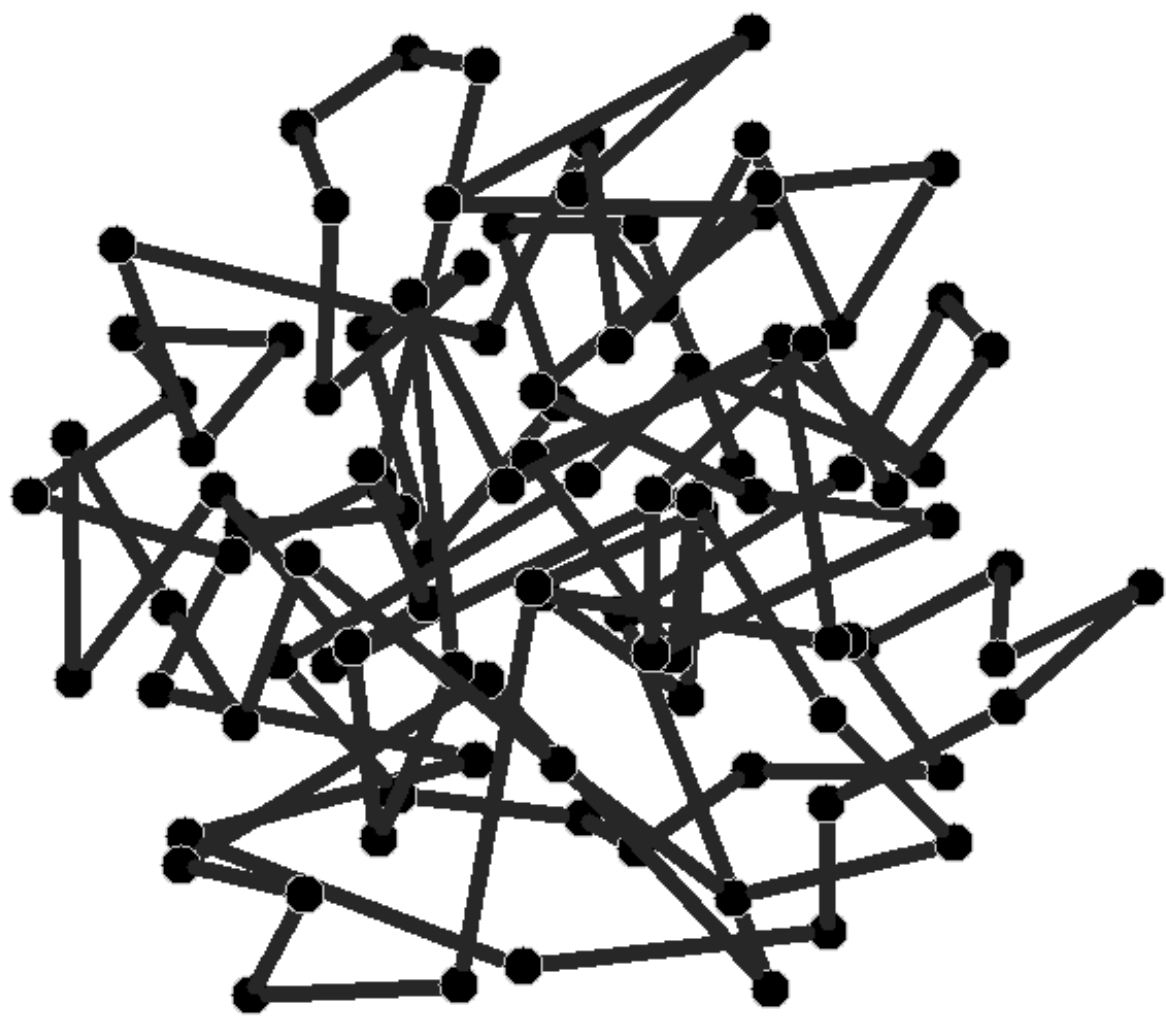
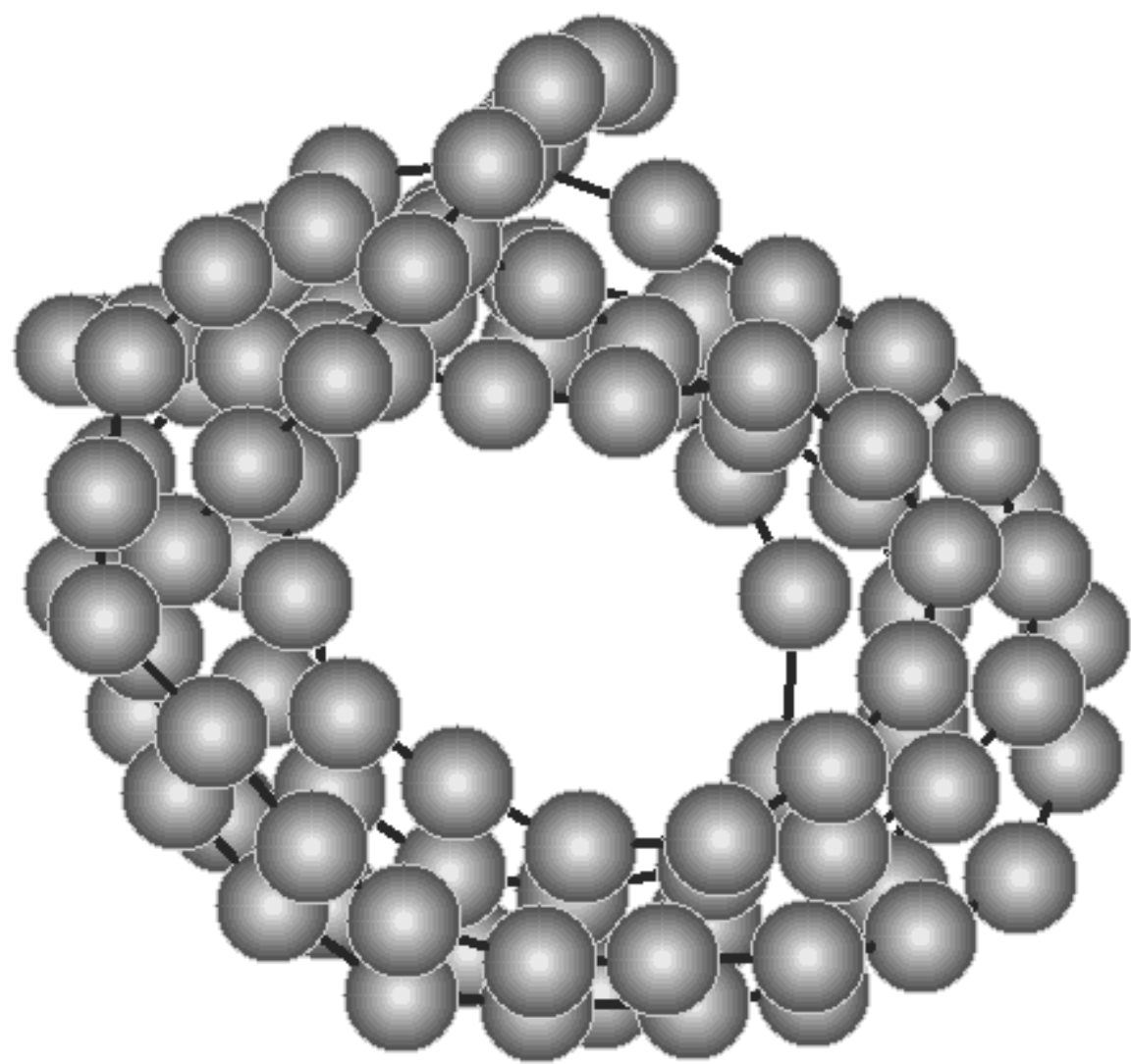
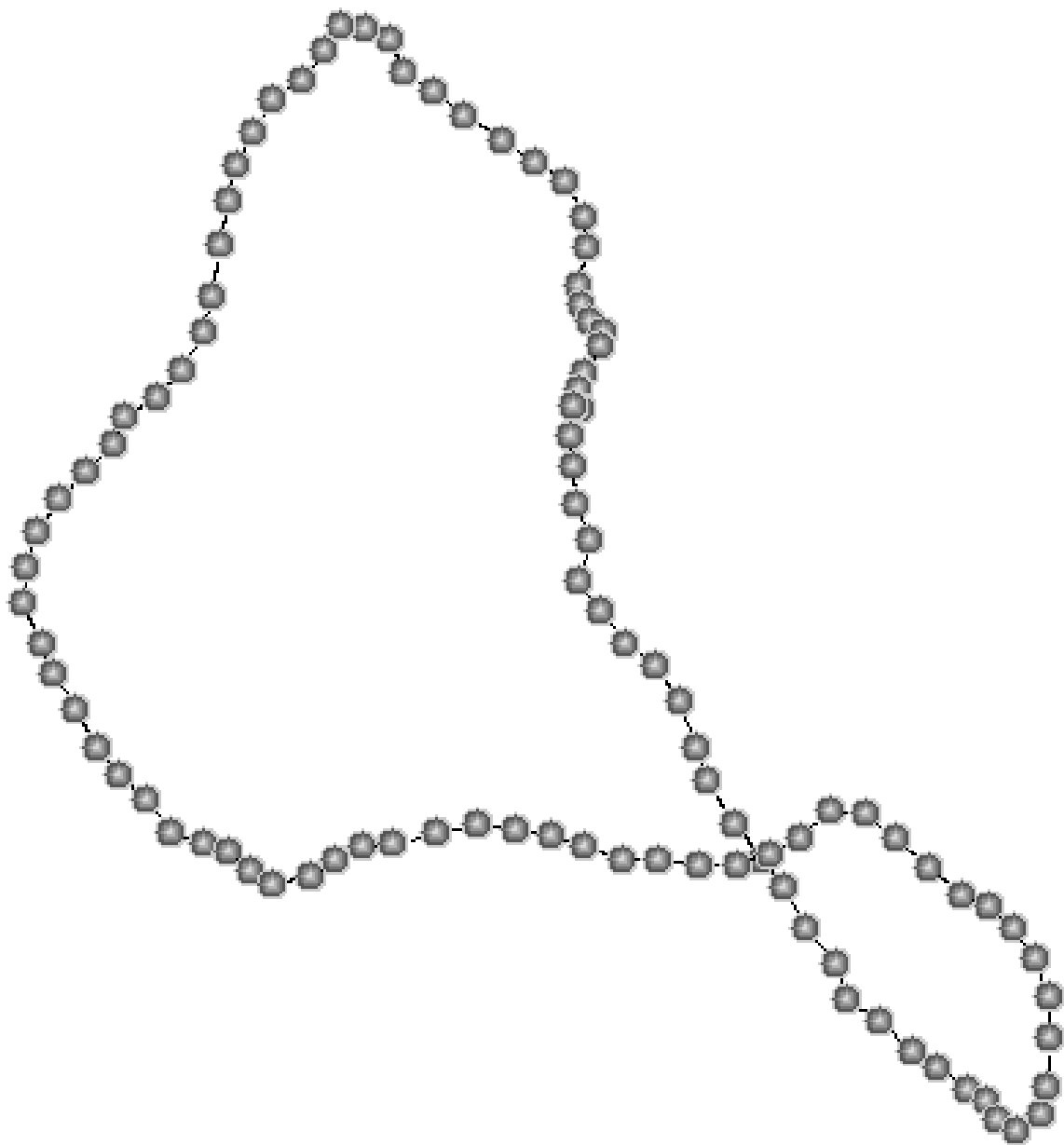
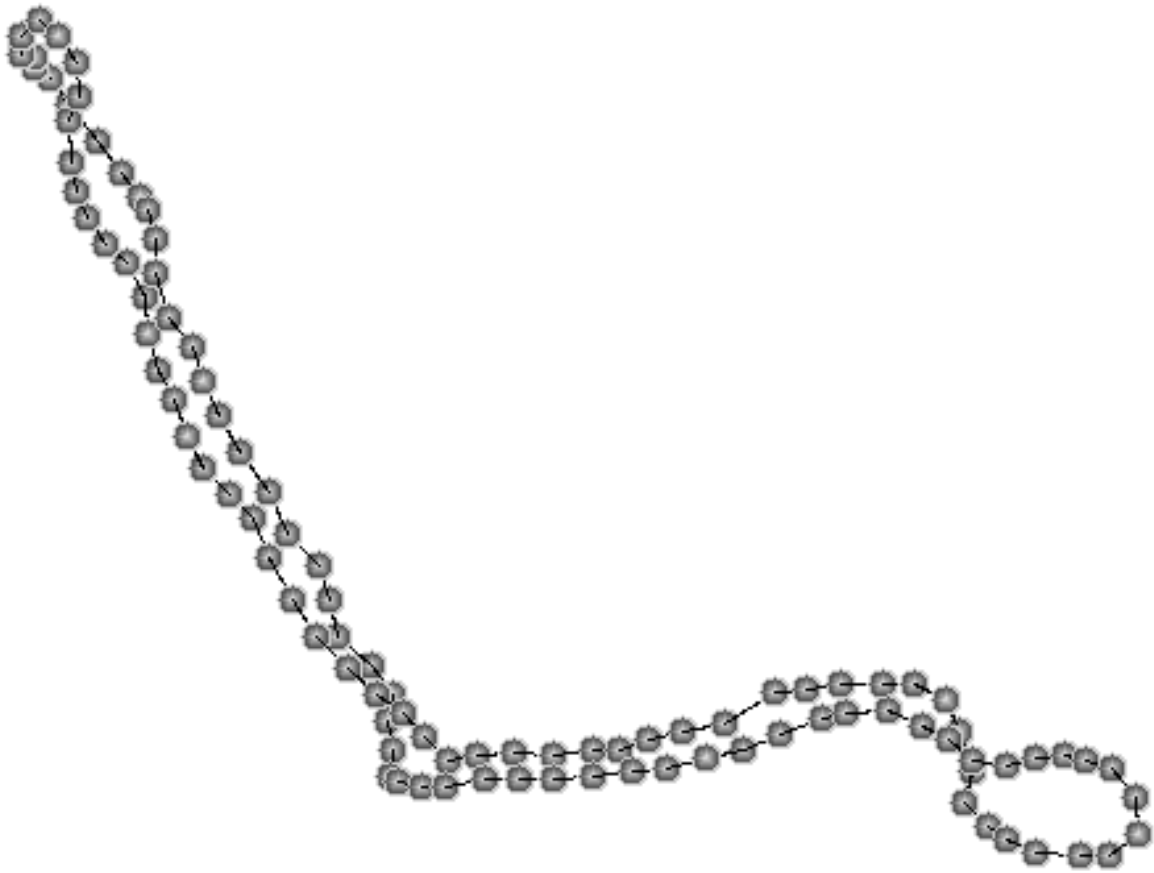
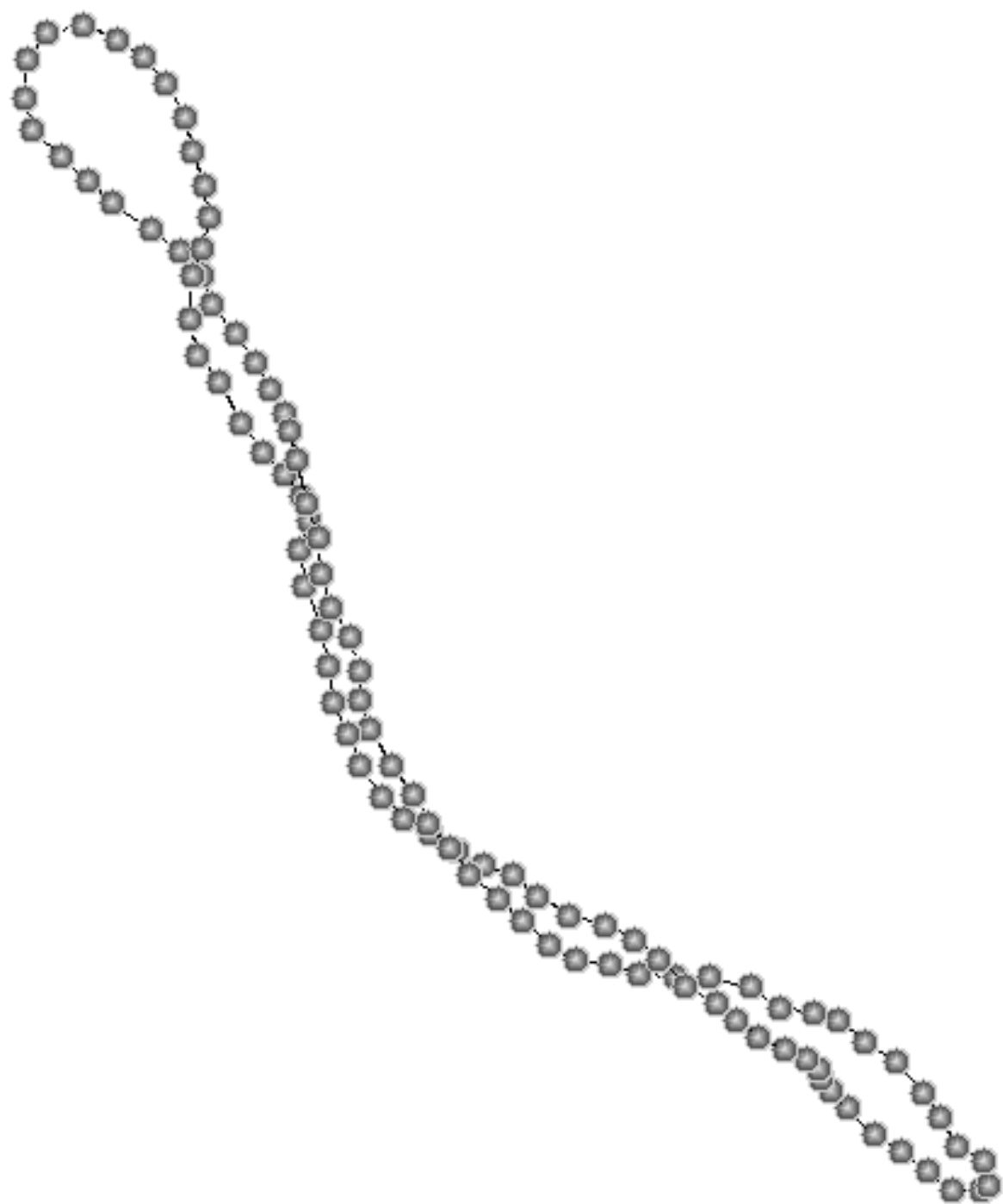
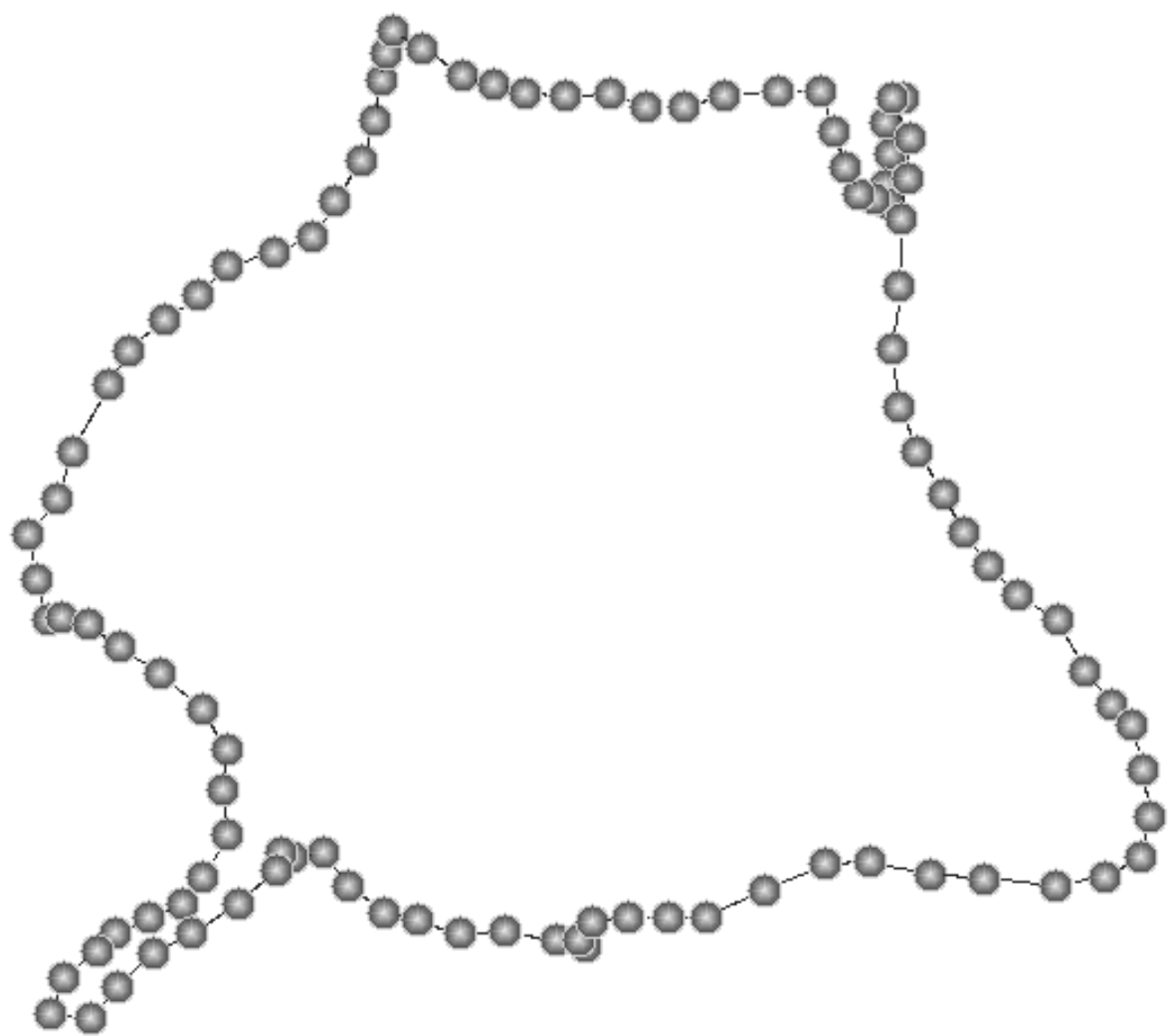
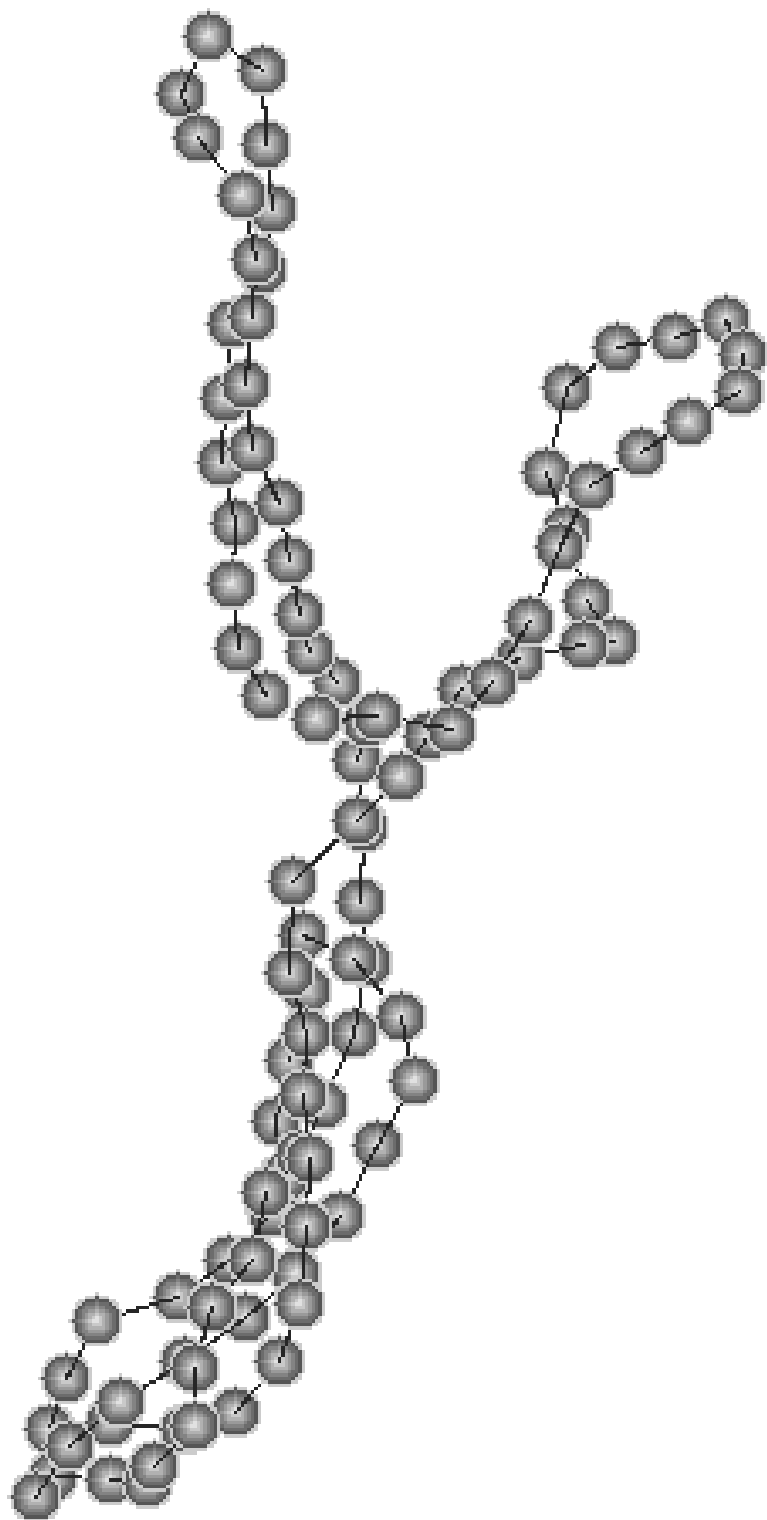
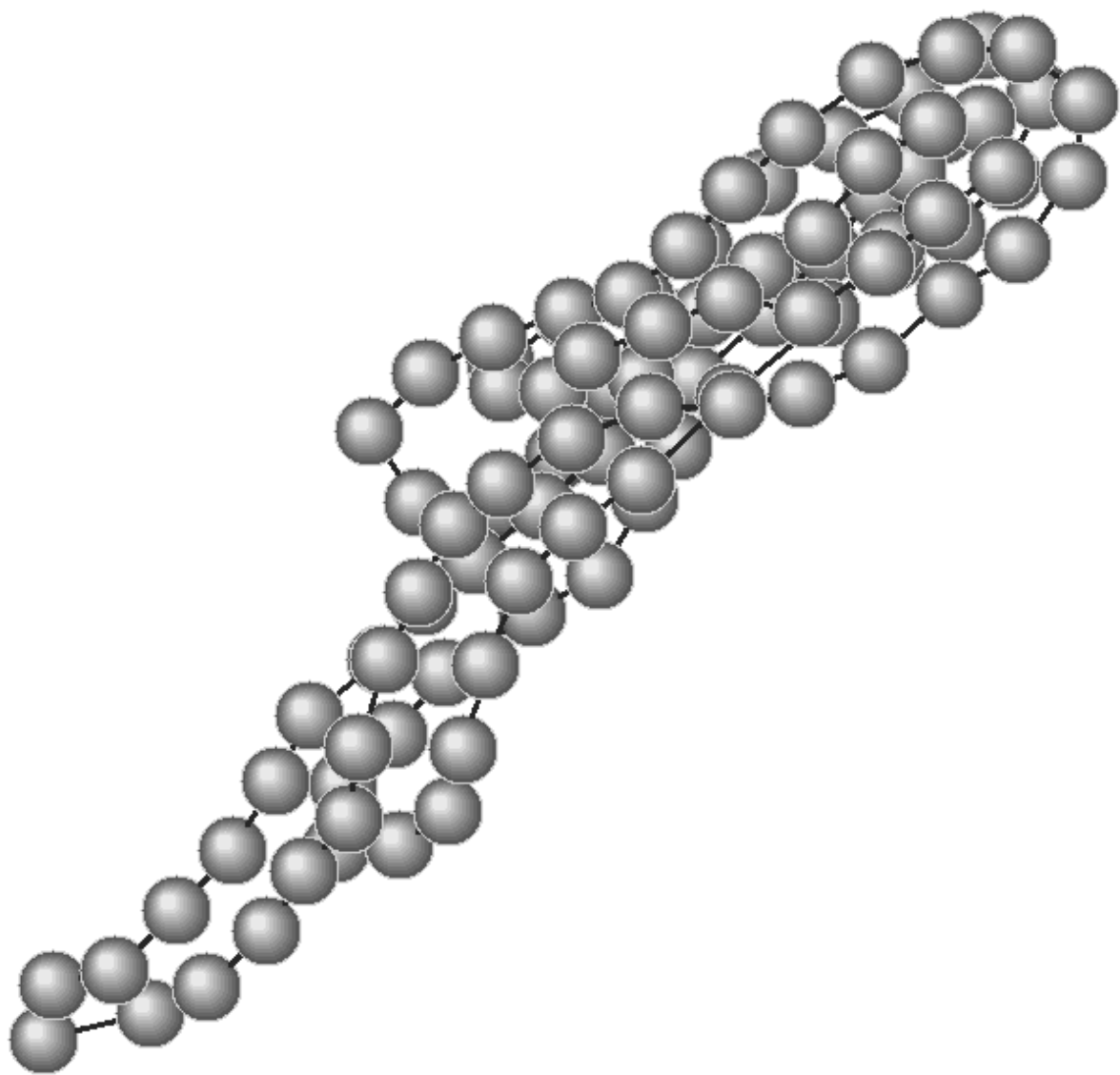
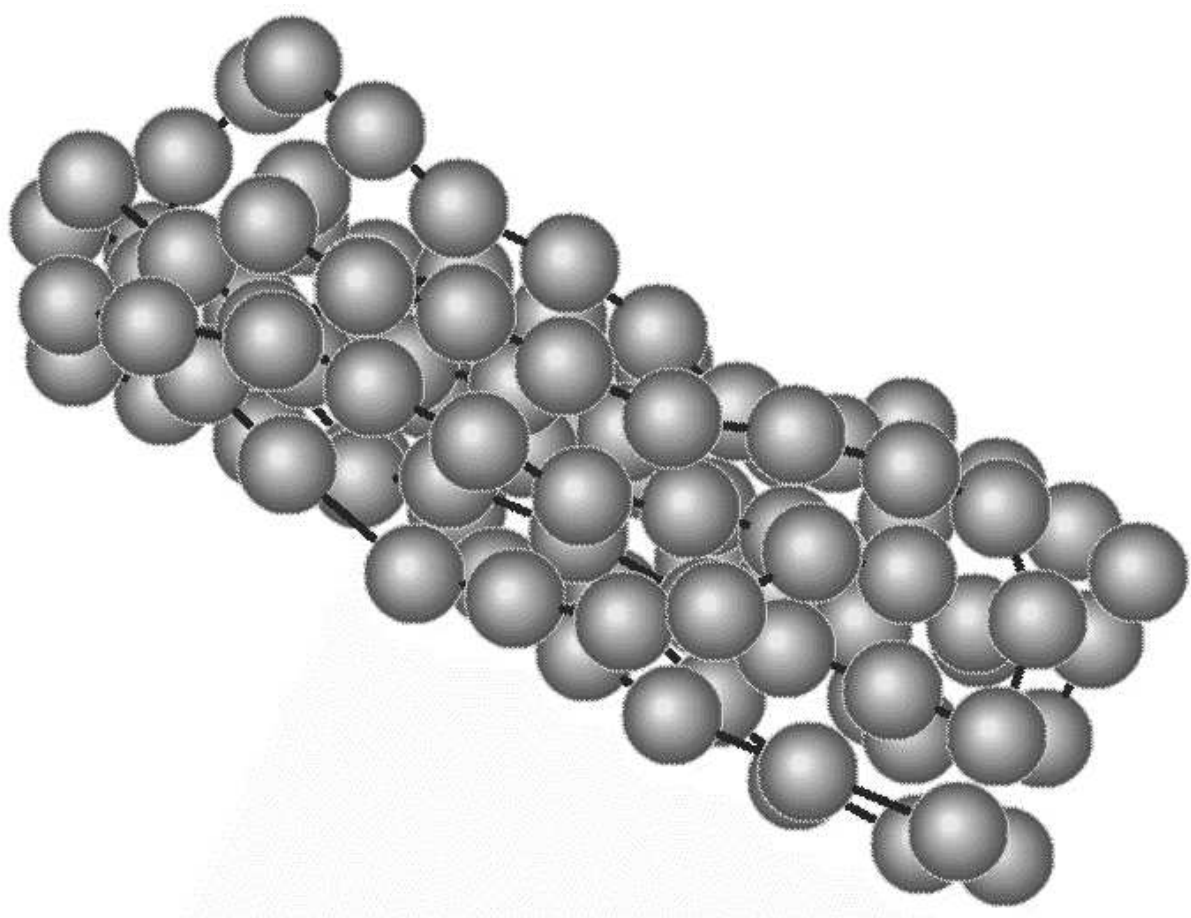
\end{document}